\documentclass[11pt]{article}

\usepackage[T1]{fontenc} 
\usepackage[utf8]{inputenc}

\usepackage{amsmath}
\usepackage[normalem]{ulem}
\usepackage{footnote}
\usepackage{amssymb}
\usepackage{hyperref}
\usepackage{graphicx}
\usepackage{wrapfig}
\usepackage{color}
\usepackage[dvipsnames,table]{xcolor}
\usepackage{epigraph}

\usepackage[square, comma, sort&compress]{natbib}
\bibpunct{(}{)}{;}{a}{}{,} % to follow the A&A style
\fontfamily{lmr}\selectfont
\newcommand*{\myfont}{\fontfamily{lmtt}\selectfont}

\newcommand{\tightitemsw}{%%
  \setlength{\topsep}{0pt}%%
  \setlength{\itemsep}{4pt}%%
  \setlength{\parsep}{0pt}%%
  \setlength{\parskip}{0pt}%%
}

\newcommand{\tightitemz}{%%
  \setlength{\topsep}{0pt}%%
  \setlength{\itemsep}{6pt}%%
  \setlength{\parsep}{0pt}%%
  \setlength{\parskip}{0pt}%%
}

\newlength{\mylen}
\setbox1=\hbox{$\bullet$}\setbox2=\hbox{\tiny$\bullet$}
\newlength\myq
\hypersetup{
  colorlinks   = true,     %Colours links instead of ugly boxes
  urlcolor     = NavyBlue, %Colour for external hyperlinks
  linkcolor    = NavyBlue, %Colour of internal links
  citecolor   =  NavyBlue  %Colour of citations
}

\usepackage{fancyhdr}
\makeatletter
\renewcommand\@seccntformat[1]{\color{CornflowerBlue} {\csname the#1\endcsname}\hspace{0.5em}}
\makeatother

\newcommand{\aap}{\textit{A\&A}}

\newcommand{\apj}{\textit{ApJ}}
\newcommand{\apjl}{\textit{ApJ}}

\newcommand{\aj}{\textit{AJ}}

\newcommand{\pasp}{\textit{PASP}}
\newcommand{\mnras}{\textit{MNRAS}}

\newcommand{\apjs}{\textit{ApJS}}

\newcounter{bibpage}

\begin{document}

%% title page
\begin{titlepage}
\hbox{ }
\vspace{-1.3cm}
\centering
{\huge \textcolor{CornflowerBlue}{Digital Infrastructure in Astrophysics} \\}
\vspace{0.5cm}
{\Large Report for the Ford and Sloan Foundation's \\ Digital Infrastructure Research Program \\ January 2020  \par \vspace{0.25cm}}
\vspace{0.5cm}
%\vfill
\begin{tabular}{ll}
Frank Timmes  & School of Earth and Space Exploration \\
              & Arizona State University, Tempe, AZ 85287 \\
              & \url{http://cococubed.asu.edu}  \\
Rich Townsend & Department of Astronomy \\
              & University of Wisconsin–Madison, Madison, WI 53706 \\
              & \url{http://www.astro.wisc.edu/~townsend/} \\
Lars Bildsten & Kavli Institute for Theoretical Physics \\
              & University of California, Santa Barbara, CA 93106 \\
              & \url{https://www.kitp.ucsb.edu}
\end{tabular}
\par
\vspace{0.5cm}
\parbox{\linewidth}{
with input 
(recorded presentations at \url{http://online.kitp.ucsb.edu/online/odia19/})
from \\
Alice Allen (University of Maryland),
Federica Bianco (University of Delaware), 
Duncan Brown (Syracuse University), 
Joel Brownstein (University of Utah), 
Kelle Cruz (Hunter College), 
Gwendolyn Eadie (University of Toronto), 
Daniel Foreman-Mackey (Flatiron Institute), 
Andrew MacFadyen (New York University),
Philipp M\"{o}sta (University of Amsterdam), 
Arfon Smith (Journal of Open Source Software and Space Telescope Science Institute), 
Jim Stone (Princeton University),
Rachel Street (Las Cumbres Observatory), and 
Matt Turk (University of Illinois Urbana-Champaign).
\\ }
%\par
\vspace{0.0cm}
\includegraphics[width=0.75\paperwidth,keepaspectratio]{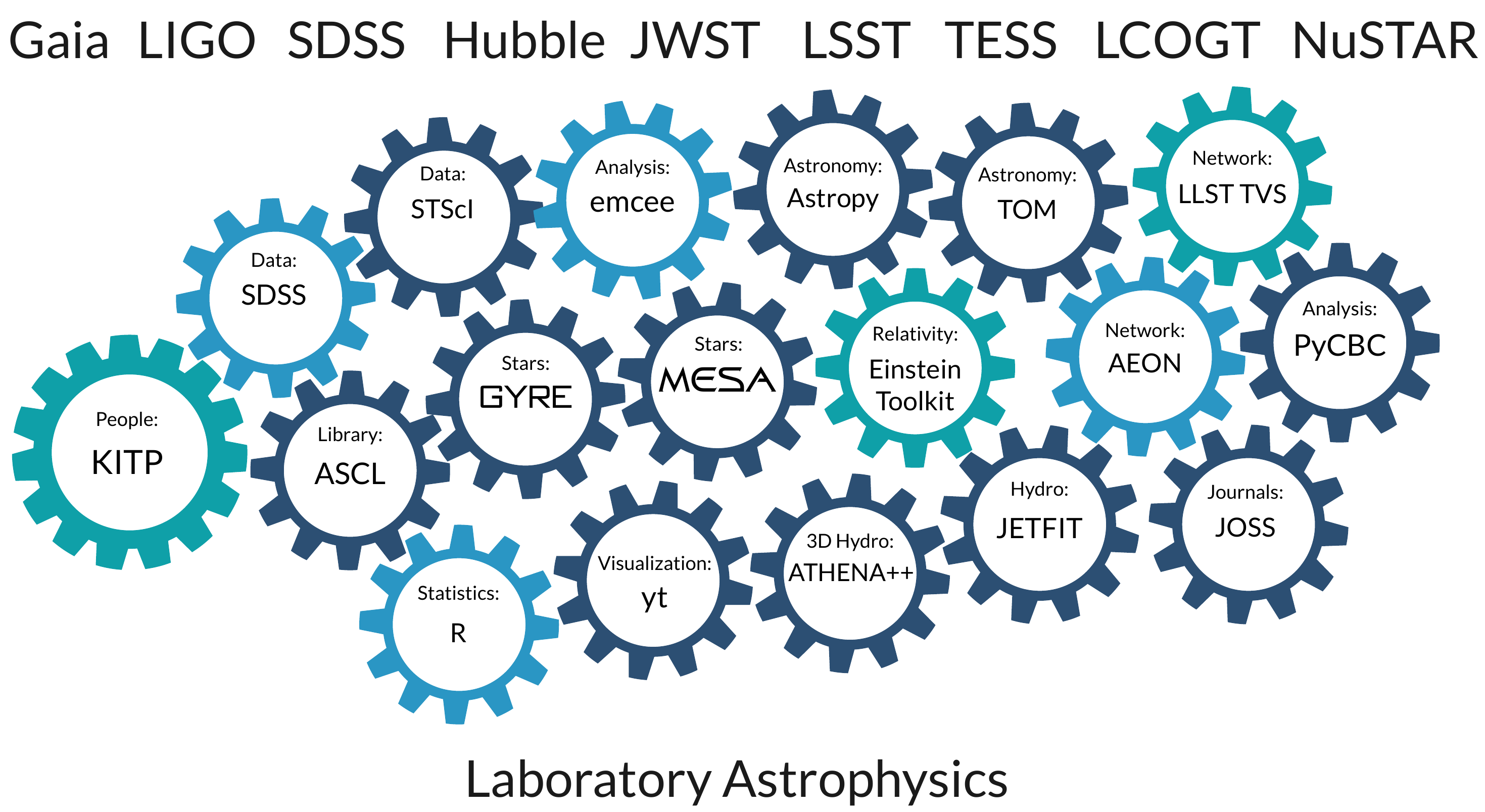}%
\vspace{0.3cm}
{\itshape This report is based on activities supported by the Ford Foundation and Sloan Foundation.
          Any opinions, findings, and conclusions or recommendations 
          expressed in this material are those of the author(s) and do not necessarily reflect the views 
          of the Ford or Sloan Foundations.\par\vspace{1cm}}
\vfill
{\par}
\end{titlepage}

% generate the table of contents
\setcounter{page}{0}
\begingroup   
\color{black} 
\tableofcontents
\endgroup   
%{
%  \hypersetup{linkcolor=black,citecolor=black}
%  \tableofcontents
%}
\thispagestyle{empty}
\clearpage

% start the page numbering
\setcounter{page}{1}

\section { {\bf \Large \textcolor{CornflowerBlue}{Executive Summary} } }\label{sec:executive}

Astronomy, as a field, has long encouraged the development of free, open digital infrastructure
\citep[e.g.,][]{national-research-council_2010_aa,national-research-council_2011_aa}.
Examples range from simple scripts that enable individual scientific
research, through software instruments for entire communities, to data
reduction pipelines for telescope operations at national facilities.
As with the digital infrastructure of our larger society today
\citep[e.g.,][]{eghbal_2016_aa}, nearly all astronomical research
relies on free, open source software (FOSS) written and maintained
by a small number of developers. 
And like the physical infrastructure of roads or bridges, digital
infrastructure needs regular upkeep and maintenance
\citep[e.g.,][]{eghbal_2016_aa}.  In astronomy, financial support for
maintaining existing digital infrastructure is generally much harder
to secure than funding for developing new digital
infrastructures that promise new science.  Sustaining 
astronomy's digital infrastructure is a new topic for many, the
sustainability challenges are not always widely known,
and sometimes even formulating answerable questions can be formidable:

\begin{itemize}\tightitemz
\item What is the relationship between money and sustainability for
community-driven, open-knowledge software instruments that enable
transformative research in stellar astrophysics? 
\item At what points in a software instrument’s lifecycle does an injection of financial resources help or hurt? 
\item Are science driven software instruments sustainable for the long term, say the next 40 years?
\end{itemize}\tightitemz

\vspace{-0.05in}
These questions are relevant from a science perspective because over
the next decade astronomy will probe the rich stellar
astrophysics of transient phenomena in the sky, including
gravitational waves from the mergers of neutron stars and black holes,
light curves and spectra from core-collapse supernovae, and the
oscillation modes of stars. {\it Laser Interferometer
Gravitational-Wave Observatory} (LIGO) and VIRGO have demonstrated the existence of
binary stellar-mass black hole systems and continue to monitor the
sky for gravitational waves from compact binary inspirals and
asymmetrical exploding massive stars. Advances in detector technology,
computer processing power, network bandwidth, software development tools, 
and data storage capability have enabled new sky surveys, such as the 
{\it Sloan Digital Sky Survey}, to create the most detailed three-dimensional 
maps of the Universe ever made.
The stellar censuses of {\it Gaia} Data Release 2, containing about one billion
stars, will provide the observational data to
tackle a range of questions related to the origin, structure, and
evolutionary history of stars in the Milky Way. This ongoing explosion
of activity powers theoretical and computational developments, in
particular the evolution of community-driven digital infrastructures for
research and education. The scientific potential of these new
observation capabilities will be unlocked largely through the 
efforts of developers and users in FOSS communities.

\vspace{0.1in}
These questions are also central from a digital infrastructure
perspective because software is an integral enabler of observation,
experiment, theory, and computation and a primary modality for
realizing the discoveries and innovations.  Answers to these questions
are also relevant from a potential funder's perspective because they
fill a knowledge gap about FOSS digital infrastructures within
domain-specific disciplines.  To date, to our knowledge, there is
little literature on the questions asked above for the domain-specific
discipline of stellar astrophysics.  Individual software instrument
project principals may know their user, bibliometric, and funding
profile history.  However, this information is usually not openly
shared or aggregated across a representative sample.  Such aggregated
information is likely to be of interest to potential funders who may
seek an informed, holistic and effective strategy to any
investment. Addressing this knowledge gap within the domain-specific
discipline of stellar astrophysics is the topic of this report.

\section { {\bf \Large \textcolor{CornflowerBlue}{Open Digital Infrastructure in Astrophysics} } }\label{sec:open}

The ``Open Digital Infrastructure in Astrophysics'' workshop was held
June 4 - 5, 2019 at the Kavli Institute for Theoretical Physics (KITP)
at UC Santa Barbara.  The workshop’s intent was to highlight
open-knowledge digital infrastructures, software instrument
communities, metrics for success, developer models, diversity efforts,
funding profiles, and sustainability plans.  The workshop strove to attain five
strategic objectives that align with the Ford and Sloan Foundation's
Digital Infrastructure Research program's broader goals:
\begin{itemize}\tightitemsw
\item Serve as a focused forum for the principals of free and open source software projects 
within the fields of astronomy and astrophysics to share knowledge with each other;
\item Explore innovative topics emerging within their respective software communities;
\item Discuss emerging best practices across the software projects;
\item Stimulate thinking on new ways of achieving long-term software sustainability;
\item Share the aggregate information and experiences in a workshop report.
\end{itemize}\tightitemsw

The 2-day workshop was held while the KITP programs 
``Better Stars, Better Planets: Exploiting the Stellar-Exoplanetary Synergy'' and
``The New Era of Gravitational-Wave Physics and Astrophysics'' were in session. 
Participants in these two programs were encouraged to attend the workshop, in part
by pausing their regular program schedules while the workshop was in session.

\section{ {\bf \Large  \textcolor{CornflowerBlue}{Vision and Execution of the Workshop}} }\label{sec:plan}

The Project Investigators (Timmes, Bildsten, and Townsend) discussed the vision, logistics, and execution 
of the workshop about once a week on average between February 2019 and June 2019.
The vision was to host speakers who spanned a diversity of FOSS projects
in stellar astrophysics:

\begin{itemize}\tightitemz
\item from single developers, through small teams, to large distributed alliance models;
\item from user communities of a few, through hunderds, to thousands;
\item from funding models of volunteers, though science/software grants, to Foundations.
\end{itemize}\tightitemz

By March 2019 the PIs converged on a list of potential speakers to ask, and by April 2019 the invited speaker list was finalized. 
The workshop website, \url{http://cococubed.asu.edu/digital_infrastructure_astronomy}, went online in May 2019.
It currently aggregates links to 

\begin{itemize}\tightitemz
\item the recorded presentations, \url{http://online.kitp.ucsb.edu/online/odia19/};
\item live tweets made during the workshop, \url{https://twitter.com/hashtag/OpenAstroInfra};
\item the FOSS projects and communities represented at the workshop;
\item efforts to conceptualize US Research Software Sustainability Institute\\
\url{http://urssi.us/};
\item posts on the fundamentals of software sustainability\\
\url{https://danielskatzblog.wordpress.com/2018/09/26/fundamentals-of-software-sustainability/};
\item a guide to sustainability models for research software projects\\
\url{https://github.com/danielskatz/sustaining-research-projects}.
\end{itemize}\tightitemz

\begin{table}[!htb]
\caption{Agenda for the Open Digital Infrastructure in Astrophysics Workshop}\label{tab:agenda}
\vspace{0.15in}
\begin{tabular}{lllll} 
\multicolumn{5}{l}{\textcolor{NavyBlue}{{\bf Tuesday June 4, 2019}}} \\
Time \cellcolor[gray]{0.9} & \cellcolor[gray]{0.9}   & Representative  \cellcolor[gray]{0.9} & Institution  \cellcolor[gray]{0.9}   & Title  \cellcolor[gray]{0.9} \\
9:00  & am  & Rich Townsend         & UW Madison         & GYRE\\
9:45  & am  & Arfon Smith           & Space Telescope    & JOSS and STScI data \\
10:30 & am  & Coffee                &                    & \\
11:00 & am  & Daniel Foreman-Mackey & Flatiron Institute & emcee \\
11:45 & am  & Federica Bianco       & Univ. of Delaware  & LSST Transients \\
12:30 & pm  & Lunch                 &                    & \\
2:00  & pm  & Matt Turk             & Univ. of Illinois  & yt \\
2:45  & pm  & Gwendolyn Eadie       & Univ. of Toronto   & R astrostatistics \\
3:30  & pm  & Break                 &                    & \\
4:00  & pm  & Joel Brownstein       & Univ. of Utah      & SDSS Data Infrastructure \\
4:45  & pm  & Rachel Street         & Las Cumbres Obs.   & TOM Toolkit / AEON Network \\
5:30  & pm  & Workshop Dinner       &                    & \\
      &     &                       &                    & \\
\multicolumn{5}{l}{\textcolor{NavyBlue}{{\bf Wednesday June 5, 2019}}} \\
Time \cellcolor[gray]{0.9} & \cellcolor[gray]{0.9}   & Representative  \cellcolor[gray]{0.9} & Institution  \cellcolor[gray]{0.9}   & Title  \cellcolor[gray]{0.9} \\
9:00  & am & Kelle Cruz         & Hunter College     & Astropy\\
9:45  & am & Jim Stone          & Princeton          & ATHENA++ \\
10:30 & am & Coffee             &                    & \\
11:00 & am & Philipp M{\"o}sta  & Univ. of Amsterdam & Einstein Toolkit \\
11:45 & am & Duncan Brown       & Syracuse Univ.     & PyCBC \\
12:30 & pm & Lunch              &                    & \\
2:00  & pm & Andrew MacFadyen   & New York Univ.     & JETFIT \\
2:45  & pm & Alice Allen        & Univ. of Maryland  & Astrophysics Source Code Library \\
3:30  & pm & Break              &                    & \\
4:00  & pm & Frank Timmes       & Arizona State Univ.  & MESA \\
4:45  & pm & Everyone           &                    & Open Discussion
\end{tabular}
\end{table}

\vspace{0.15in}
\noindent
%{\bf \Large  \textcolor{CornflowerBlue}{3. Workshop Agenda and Focus Questions}}
\section{ {\bf \Large  \textcolor{CornflowerBlue}{Workshop Agenda and Focus Questions}} }\label{sec:agenda}

Table \ref{tab:agenda} lists the invited speakers, their home
institution, and their digital infrastructure project. 
The PIs encouraged the invited speakers to consider presenting quantitative and
qualitative information, as appropriate, on the following 18 high-value questions and topics. 
In these questions, the word ``instrument'' was intended to be broadly interpreted to mean 
an open-source software instrument, an open data repository, or an open library.

\begin{itemize}\tightitemsw
\item What does your FOSS instrument do?
\item Who is your instrument's community?
\item How many users does your instrument have, and how is this metric tracked?
\item How is communication with your user base conducted, and at what frequency?
\item What efforts are made to expand the user base?
\item What are the longitudinal bibliometrics (citations, citations to citations, rankings, honors)?
\item How often is an article published/distributed that describes new capabilities?
\item What offshoots or spinoffs from the software instrument have arisen?
\item Who is the developer community?
\item What is the developer model?
\item How are new developers attracted?
\item How are developers retained?
\item What efforts are made to build a diverse developer base?
\item What is the past funding profile for your instrument?
\item What new funding efforts are underway?
\item Offer a definition of ``sustainability'' for your instrument.
\item What efforts are being made to reach this definition of sustainability?
\item Will your instrument be widely used by your community in 10, 20, 40 years?
\end{itemize}

\section{ {\bf \Large  \textcolor{CornflowerBlue}{The Digital Infrastructure Projects Considered}} }\label{sec:projects}

In this section we present data on the digital infrastructure projects studied.

\subsection{ {\bf \large \textcolor{CornflowerBlue}{GYRE - Rich Townsend}} }\label{sec:gyre}

Like a sound wave resonating in an organ pipe, sound waves can
resonate inside a star. By measuring these wave frequencies, we can
learn about a star’s internal structure. This field, asteroseismology,
has been reinvigorated in recent years thanks to the wealth of new
observational data provided by NASA space-based instruments such as
{\it Kepler} and now {\it TESS}. Interpreting these new observations
requires the astronomer's analogue to the telescope: a stellar
oscillation software instrument which calculates the frequency
spectrum of an arbitrary input stellar model.

\vspace{0.1in}
The free and open GYRE source code solves the system of equations
governing small periodic perturbations to an equilibrium stellar
state. GYRE is built on a robust and accurate numerical scheme,
and makes efficient use of multiple cores and/or cluster nodes.
The suite is written in Fortran 2008, with an object oriented architecture
that is simple, clean, modular and adherent to modern coding best
practices (\url{https://bitbucket.org/rhdtownsend/gyre/wiki/Home}).  

\vspace{0.1in}
GYRE's target community is stellar asteroseismologists.  
GYRE has currently attracted over 200 users world-wide.
GYRE  innovates by updating its community with an instrument paper describing major
new software and science capabilities about every five  years
\citep{townsend_2013_aa,townsend_2018_aa}. 
These two instrument papers have been cited $\simeq$ 150 times
and have a current citation rate of $\simeq$ 65/year.  The articles
that cite GYRE have themselves generated $\simeq$ 3,271 citations,
yielding a radius-of-influence of $\simeq$ 3,271/150 $\simeq$ 21, which
suggests that GYRE helps generate articles that the broader
astronomy community values.
GYRE provides a portal for its community to openly share knowledge
(\url{http://www.astro.wisc.edu/~townsend/gyre-forums/}), and offers
$\simeq$ 750 archived and searchable posts of community discussions
about stellar oscillations.  

\vspace{0.1in}
GYRE is maintained by two developers and five contributors
who directly support a community of $\simeq$ 110 users and indirectly
a community of $\simeq$ 300 users uses through GYRE's integration with MESA
(see Section \ref{sec:mesa}). The combination of these users downloaded 
the latest release of GYRE $\simeq$ 400 times.
Development of GYRE and its community have been supported by a 
National Science Foundation (NSF) science award between 2009-2002 for \$428K,
NSF digital infrastructure award between 2014-2016 for \$76K,
NSF digital infrastructure award between 2017-2021 for \$683K,
and a NSF science award between 2017-2020 for \$317K.

\subsection{ {\bf \large \textcolor{CornflowerBlue}{Space Telescope Science Institute (STScI) Data - Arfon Smith}} }\label{sec:stsci}

The STScI helps humanity explore the universe with advanced space
telescopes and ever-growing data archives.  The Data Science Mission
Office (DSMO) within STScI is responsible for maximizing the
scientific returns from archives containing astronomical observations
from 17 space-based missions and ground-based observatories.  The
scope of the DSMO includes Data Science (statistics, machine learning,
algorithms, analytics and data mining) and all aspects of data management.

\vspace{0.1in}
The James Webb Space Telescope (JWST) is a space telescope that is
planned to be the successor to the Hubble Space Telescope (HST).  The
main focus of JWST data analysis tool development is to advance a set
of high-quality FOSS libraries for working with JWST data, leveraging
FOSS technologies where possible, In addition, the aim is to provide
a set of high-performance visualization tools for exploring JWST data
products, and provide a comprehensive suite of reproducible analysis
captured as Juptyer notebooks.

\vspace{0.1in}
The vision for these tools is to ensure that astronomers are equipped
with software to analyze and interpret JWST efficiently from the start
of the mission; reduce the necessity for astronomers to write
data-analysis software, but make it easy to do so when necessary;
provide a clear mechanism for astronomers to share expertise with each
other and make it easy to share and re-use code; allow the community
to easily contribute to and expand the capabilities of existing tools;
improve repeatability and reliability of scientific results; and
provide a clear offramp for those astronomers not wishing to make use
of the tools developed by the mission.

\vspace{0.1in}
The guiding principles for the JWST data analysis tools are that they
should be FOSS, easy to install, well documented, easy to extend; able
to support scripting and graphical user interfaces; built on stable,
widely adopted languages and libraries; leverage existing source codes
and algorithms, including some developed outside astronomy; allow for
easy contributions by non-STScI staff; provide a consistent, coherent
experience across all STScI-developed data analysis tools, regardless
of scientific use case.  Some of the FOSS developed to date 
include JWST-specific libraries, Astropy contributions (see Section
\ref{sec:astropy}), and contributions from the STScI community
software initiative.  The development model is internal development
and See You On GitHub.

\vspace{0.1in}
The Mikulski Archive for Space Telescopes (MAST) aims to maximize the
scientific accessibility and productivity of astronomical data.  Due
to supporting several missions and observatories, the data products are
heterogeneous and complex. MAST currently holds about 3 petabyte of
data products, which is currently downloaded about once every 18
months.  HST data currently generate $\simeq$ 800 articles/year, with
$\simeq$ 300 articles/year from archival data. The earliest Hubble
observations still yield publishable results almost three decades later.

\vspace{0.1in}
The Data Science Mission Office's community is the world-wide
astronomical community.  Communication with users primarily occur via
newsletters, mailing lists, and conferences.  Long-term sustainability
is the astronomy community being able to access (and interpret) data for
decades to come. Funding for STScI comes from government contracts, mainly through NASA.

\subsection{ {\bf \large \textcolor{CornflowerBlue}{Journal of Open Source Software (JOSS) - Arfon Smith}} }\label{sec:joss}

JOSS is a developer-friendly, open access journal for research
software packages.  It is committed to publishing quality research
software with zero processing charges or subscription fees to authors.
JOSS is an academic journal (ISSN 2475-9066) with a formal, public,
iterative peer review process that is designed to improve the quality
of the software submitted (\url{https://joss.theoj.org}).

\vspace{0.1in}
JOSS publishes articles about research software that: solves complex
modeling problems in a scientific context (physics, mathematics,
biology, medicine, social science, neuroscience, engineering);
supports the functioning of research instruments or the execution of
research experiments; extracts knowledge from large data sets; offers
a mathematical library; or similar. JOSS submissions must be open
source i.e., have an Open Source Definition (OSD) compliant license;
have an obvious research application; be feature-complete (no
half-baked solutions) and be designed for maintainable extension (not
one-off modifications); not be minor ``utility'' packages, such as
``thin'' application programming interface (API) clients. JOSS
currently publishes $\simeq$ 30 articles/month. Each article receives
a Digital Object Identifier (DOI) and is indexed by the Astrophysics
Data System (ADS), an online database of over eight million astronomy
and physics articles from both peer reviewed and non-peer reviewed sources.

\vspace{0.1in}
JOSS's community is academics writing open source software.
Communication between editors, authors, and potential authors takes
place in the open on Twitter, a blog, and GitHub.  Offshoots include
the sister journals Journal of Open Source Education and the Julia
conference proceedings.  JOSS is mainly funded by a grant from the
Sloan Foundation and a volunteer editorial board.  JOSS probably 
probably/hopefully not be needed in 5, 10, or 20 years as long-term sustainability
is the academic credit model moving away from its exclusive focus on
academic articles and citations.

\vspace{0.1in}
The American Astronomical Society (AAS) Journals are arguably the
top-tier forum for publishing significant new research in astronomy
and astrophysics (\url{https://journals.aas.org}).  In 2016 the AAS
Journals adopted a policy that reflects the importance of software to
the astronomical community, and the need for clear communication about
such software which ensures that credit is appropriately given to its
authors. The policy provides clear guidelines for citing software in
all manuscripts, and supports the publication of descriptive articles
about software relevant to research in astronomy and astrophysics
(\url{https://journals.aas.org/policy-statement-on-software/}).

\vspace{0.1in}
Since adopting this software policy an increasing number of software 
publications have appeared. However, many authors and readers would like the
AAS Journals to provide a more detailed review of the software itself. 
Such a review might provide useful feedback to authors and provide readers with 
further assurance that software packages are of a ``gold standard'' quality.

\vspace{0.1in}
In order to provide this service, the AAS Journals partnered with the
JOSS in 2017.  Authors submitting articles to the AAS Journals
describing software published with a OSD-compliant license may choose
to have their software reviewed in parallel at JOSS. Upon completion
of both review processes, the AAS Journals indicate on the published
article that the software has also been reviewed. The AAS Journals are
paying a modest fee to JOSS for this service ($\simeq$ \$50/article),
which helps support their operations and which is not passed on to
authors, referees, or editors.  The increasing number of authors who
use this partnership between JOSS and the AAS Journals suggests the
FOSS astrophysics community finds it a useful service.

\subsection{ {\bf \large \textcolor{CornflowerBlue}{emcee - Daniel Foreman-Mackey}} }\label{sec:emcee}

Probabilistic data analysis, which describes degrees of logical
implication or subjective certainty, has transformed much of
scientific research in the past decade.  Many of the most significant
gains have come from numerical methods for approximate inference,
especially Markov chain Monte Carlo (MCMC).  For example, many
problems in astrophysics have directly benefited from MCMC because the
models are often expensive to compute, there are many free parameters,
and the observations usually have low signal-to-noise ratios
\citep{foreman-mackey_2013_aa}.

\vspace{0.1in}
The emcee package is a Python implementation of a novel MCMC
sampler that enables efficient Bayesian inference
(\url{https://github.com/dfm/emcee}).  Bayesian inference, as one
probabilistic data analysis method, provides a powerful general basis
for data analysis. The algorithm behind emcee has several advantages
over traditional MCMC sampling methods and it has excellent
performance as measured by the autocorrelation time (or function calls
per independent sample).

\vspace{0.1in}
The emcee package has been widely applied to probabilistic modeling
problems in astrophysics with some applications in other fields.  For
example, the instrument paper describing emcee is currently in the 
\href{https://tinyurl.com/mesatwo2013}{Top 3}
most cited articles in astronomy that were published in 2013 ($\simeq$
3100 citations, citation rate of $\simeq$~1100/year).  The articles
that cite emcee have themselves generated $\simeq$ 120,000 citations,
yielding a large radius-of-influence of $\simeq$ 40, which strongly
indicates the emcee package helps generate science articles that
the broader astronomy community values.

\vspace{0.1in}
emcee is currently maintained mostly by one developer with smaller
contributions from $\simeq$ 8 contributors.  Version 3.0 of emcee, to
be released in 2020, is the first major release in about 6 years.  It
includes a full re-write of the computational backend, several
commonly requested features, and a set of new ensemble
implementations. To date, emcee has not been directly funded.
Instead, development of the source code has relied on being part 
of a funded science effort.

\subsection{ {\bf \large \textcolor{CornflowerBlue}{LSST Transients - Federica Bianco}} }\label{sec:lsst}

A transient astronomical event, often shortened to ``transient'', is an
astronomical object or phenomenon whose duration may be from seconds
to days, weeks, or even years. This is in contrast to the timescale of
the millions or billions of years during which the galaxies and their
component stars evolve.  Singularly, the term is used for intrinsic
events such as supernovae, novae, dwarf nova outbursts, gamma-ray
bursts, and tidal disruption events, as well as geometric events such
as gravitational microlensing, transits, occultations, and eclipses.

\vspace{0.1in}
The science community interested in the Large Synoptic Survey Telescope (LSST) is organized into Science
Collaborations.  Among those, the LSST Transients and Variable Stars
Collaboration focuses on the transient sky (\url{https://lsst-tvssc.github.io}), 
which is one of the 4 primary science drivers of LSST
(\url{https://www.lsst.org/science}).  The collaboration is organized
in 15 subgroups by common expertise and interests, that are
delineating a roadmap to maximize the potential of the LSST survey in
the discovery and characterization of the transient sky.

\vspace{0.1in}
The LSST data reduction pipeline will deliver an alert package for
every object in each image that exhibits a significant brightness or
positional change.  The shear scale of the survey leads to an
anticipation of tens of millions of alerts per night and hence a
computational challenge for astronomers to identify targets of
specific interest for their science.  In this context, ``broker'
refers to software which receives alert information, associates it
with other data, performs classification functions according to
numerous algorithms and criteria, and stores the information in a
database. Brokers will provide interfaces to enable users to sort
targets and alerts according to their own preferences. These users may
be individual scientists, teams, or additional brokers.
Brokers may also be accessed by the TOM
software (See section \ref{sec:toms}), which is designed to
automatically prioritize targets for a specific science goal, request
follow-up observations, track progress, and analyze the results.  To
maximize the science return from LSST, the user interfaces to these
brokers need to be carefully designed to address the needs of the
community.

\vspace{0.1in}
Financial support for the LSST's construction, operation, and its digital infrastructure,
comes from the NSF, the Department of Energy,
and private funding (e.g., Google, Simonyi Foundation) raised by the LSST Corporation, 
a non-profit 501(c)3 corporation formed in 2003.

\subsection{ {\bf \large \textcolor{CornflowerBlue}{yt - Matt Turk}} }\label{sec:yt}

Conducting and analyzing computational simulations in astrophysics has
grown increasingly sophisticated. The resultant data encompass many
types of physical models, span tens of thousands of timesteps, and
consume terabytes of storage with a variety of metadata schemes.  As a
result, analysis routines were often focused around specific problems,
and specific simulation platforms that produced duplicated efforts, or
incompletely analyzed simulations. yt was developed to alleviate these
issues \citep{turk_2011_aa,turk_2013_aa}.

\vspace{0.1in}
yt is a community-driven, permissively-licensed, FOSS Python package
for analyzing and visualizing volumetric data (\url{https://yt-project.org}).  
It supports structured, variable-resolution meshes, unstructured
meshes, and discrete or sampled data such as particles.  It is focused
on driving physically-meaningful inquiry that meet the challenges of
reproducibility, parallelization, and vast increases in data size and
complexity. yt has been extensively applied in domains such as
astrophysics, seismology, nuclear engineering, molecular dynamics, and
oceanography. The yt software instrument article is currently one of the 
\href{https://tinyurl.com/mesaone}{Top 50}
most cited articles in astronomy and astrophysics that were 
published in 2011.

\vspace{0.1in}
yt is unique in having a very large, active development team.  Over
the entire history of the project, 186 developers have contributed.
Over the past twelve months, 35 developers contributed new code to yt.
yt has one of the largest open-source teams in the world, ranking in
the top 2\% of all project teams on Open Hub.  Development of yt is
driven by a commitment to Open Science principles as manifested in
participatory development, reproducibility, documented and
approachable code, a friendly and helpful community of users and
developers, and Free and Libre Open Source Software.
yt is supported in part by the Gordon and Betty Moore Foundation
(\$1.5M and \$350K between 2014-2019) and the NSF (\$85K, \$500K and \$1.65M between years 2014). 
yt is a also fiscally-sponsored project of NumFOCUS, a 501(c)(3) public charity,
whose mission is to promote open practices in research, data, and scientific 
computing by serving as a fiscal sponsor for open source projects and 
organizing community-driven educational programs.

\subsection{ {\bf \large \textcolor{CornflowerBlue}{R Astrostatistics - Gwendolyn Eadie}} }\label{sec:r}

R is a programming language and FOSS environment for statistical
computing and interactive graphical analysis (\url{https://www.r-project.org}).  
R and its libraries implement a wide variety of statistical and
graphical techniques, including linear and nonlinear modeling,
classical statistical tests, nonparametrics, multivariate regression
and multivariate classification, time series analysis, and Bayesian
inference, and others.  R is extensible through functions and
extensions, and the R community is known for its active contributions.
Many of R's standard functions are written in R itself, which makes it
easier for users to follow the algorithmic choices made. For
computationally intensive tasks, C, C++, and Fortran packages can be
linked and called at run time.

\vspace{0.1in}
R is widely used among statisticians for developing novel statistical
software and data analysis.  Python currently dominates in astronomy
and R is less common, perhaps because most astronomers are not (yet)
usually trained in statistics.  An advantage of R over Python is
exposure to new developments in statistical methods as R is the
language of statisticians. Astrostatistics is a discipline on the rise
as it helps process the vast amount of data produced by automated
scanning of the cosmos, to characterize complex datasets, and to link
astronomical data to astrophysical theory. For example, the
Astrostatistics and Astroinformatics Portal (\url{http://asaip.psu.edu})
serves the cross-disciplinary communities of astronomers,
statisticians and computer scientists. It is intended to foster
research into advanced methodologies for astronomical research, and to
promulgate such methods into the broader astronomy community.
Packages have also been written to integrate R and Python within both the R environment and Jupyter Notebooks.

\vspace{0.1in}
R is an official part of the Free Software Foundation’s GNU project,
which is a principal source of funding for the development,
maintenance, and community-support of R.  The R Foundation is a
not-for-profit organization working to support the R project and other
innovations in statistical computing
(\url{https://www.r-project.org/foundation/}).  The R Consortium is a
group organized under an open source governance and foundation model
to support the global community of users, maintainers and
developers of R software (\url{https://www.r-consortium.org}). Its
members include leading institutions and companies dedicated to the
use, development and growth of R.  Example of its community
building efforts include supporting useR!, the annual meeting of the R
user and developer community (e.g., \url{https://user2020.r-project.org}), 
and R-Ladies, a global organization whose mission is to promote
gender diversity in the R community (\url{https://rladies.org/about-us/}).  

\vspace{0.1in}
The Comprehensive R Archive Network (CRAN,
\url{https://CRAN.R-project.org/}) is the official, regulated,
peer-reviewed repository of R.  CRAN has a set of policies on package
submissions and updates; these policies provide a standardized format
for R packages, help files, and example code, as well as required
testing and compatibility with future R distributions
(\url{https://cran.r-project.org/web/packages/policies.html}).  For
example, when a new version of R is released, if a package no longer
works then the maintainer is required to update their package. If no
update is completed within a certain amount of time, then the package
is archived.  CRAN aggregates a collection of sites which carry
identical material, consisting of the R distribution(s), the $\simeq$
14,000 contributed extensions, documentation for R, and binaries.
While R is available in Jupyter Notebooks, RStudio is the most popular
user interface for R (\url{https://rstudio.com/}), and R Shiny is a
popular tool for developing interactive figures and graphs
(\url{https://shiny.rstudio.com/}).

\subsection{ {\bf \large \textcolor{CornflowerBlue}{SDSS Digital Infrastructure - Joel Brownstein}} }\label{sec:sdss}

Astronomy sky surveys are research projects to capture uniform data
about a region of the sky. 
SDSS has a long history of operating advanced data acquisition,
innovative data management, world-class archival storage, and
scientific visualization practices. The breadth and depth of science
fostered by SDSS's community-organized approach
(\url{https://www.sdss.org}) have advanced our understanding of the
cosmos, and spurred innovative technical and analytical approaches to
challenges in systems development, data management, and pipeline
development and implementation.  SDSS has generated over 8000 refereed
articles, currently over 600 refereed articles/year, and $\simeq$
330,000 citations.  The road to success was not without lessons about
how to conduct big, collaborative, data-based scientific projects.

\vspace{0.1in}
SDSS data is accessed through the Catalog Archive Server (CAS), the
Science Archive Server (SAS), or direct data file access via globus online,
http by wget or curl \citep{weaver_2015_aa,cherinka_2019_aa}.  CAS offers
catalog level data and interactive tools to browse through SDSS images
with links to associated spectra and catalog data about the objects on
the images.  SAS is the primary interface to 
all of the original images, raw data from North and South Observatories,
and includes all of the spectra, intermediate and final results, and 
is accessed at rates exceeding 10 TB of downloads and 50 million web hits per month.
It provides tools to interactively view and download SDSS
spectra, download images of SDSS fields, and generate mosaics of those
fields. Storage at SAS crossed the 1 petabyte level in 2019.  The
final Data Release of SDSS-IV is scheduled for July 2021, and will
include all APOGEE-2, eBOSS and MaNGA spectra observed during SDSS-IV,
as well as all final data products and catalogs SDSS-V will start
observations in summer 2020, with its first data release expected two
years later. Surveys in SDSS-V include Milky Way Mapper, Local Volume
Mapper and Black Hole Mapper.  A challenge for the near-future is
developing and maintaining a sustainable and innovative science
archive.

\vspace{0.1in}
Funding for the Sloan Digital Sky Survey IV has been provided by the
Alfred P. Sloan Foundation, the U.S. Department of Energy Office of
Science, and the Participating Institutions.  The Sloan Foundation was
an early supporter of the SDSS in 1992 and has invested $\simeq$ \$60M
across 16 grants, with another \$16M grant recently approved to
continue SDSS through at least 2024
(\href{https://www.sciencephilanthropyalliance.org/science-philanthropy-success-story-the-sloan-foundations-25-year-partnership-with-the-sloan-digital-sky-survey/}{https://www.sciencephilanthropyalliance.org}).
The project is widely considered one of the big success stories in science
philanthropy. The Sloan Foundation has provided $\simeq$ 20–25\% of
the SDSS costs at each phase of its implementation. The balance has
been funded mainly by Participating Institutions that join as
members. Each member typically contributes $\simeq$ \$1 million over
five years, which allows its faculty, researchers, and students to get
early access to SDSS data. The project now has over 54 university
partners who contribute most of the funding for SDSS.

\subsection{ {\bf \large \textcolor{CornflowerBlue}{TOM Toolkit - Rachel Street}} }\label{sec:toms}

Astronomical surveys are producing ever increasing catalogs of new
discoveries at ever faster rates. Astronomers have found it necessary
to build database-driven systems -- called Target and Observation Managers
(TOMs) -- for powerful, programmable control over all aspects of astronomical observing
schedules and data products.  TOM systems offer users a powerful
way to display and interact with targets and their data through a web browser.
These systems can also submit requests
for observations directly to networked, robotic telescope facilities
to harvest the data products.  When coupled with analysis
software, TOMs are capable of conducting entirely automated
follow-up programs, including rapid responses to national and
international alerts.

\vspace{0.1in}
Building a TOM system previously required specialist expertise in database
and software development, thus generally restricting TOMs to a subset
of large projects.  The TOM Toolkit is a FOSS software package
centered around a highly flexible database, designed for astronomical
data and to be customized by the user to accommodate science-specific
parameters and data products.  Developed by professional software
engineers in collaboration with scientists at Las Cumbres Observatory
(\url{https://lco.global/tomtoolkit/}), the TOM Toolkit can be used as a
stand-alone package to build a TOM from scratch or as a library of
useful functions.  The Toolkit comes with a wide-range of functions to
support observing programs and is designed to be extendable with
community contributions.  The TOM Toolkit is designed to make it easy
to develop plugins to interface with external software and resources,
such as additional telescope facilities.
Development of the TOM Toolkit is supported the Zegar Family
Foundation (\$450K between 2018 and 2020) Heising-Simons Foundation
(\$650K between 2018 and 2020).

\subsection{ {\bf \large \textcolor{CornflowerBlue}{AEON Network - Rachel Street}} }\label{sec:aeon}

Modern astronomical surveys can deliver tens of thousands of new
discoveries every night, alerted within minutes.  Yet many will
require additional observations in order to understand the physical
phenomena and maximize the scientific return.  Observatories supporting
this critical follow-up must be capable of responding on similar
timescales and with a flexibility governed by the demands of the science.
The Astronomical Event Observatory Network (AEON, \url{http://ast.noao.edu/data/aeon}) 
is a collection of world-class telescope facilities which can be accessed on demand, 
at the touch of a button.  At the heart of the network, the Las
Cumbres Observatory is joining forces with the National Optical
Astronomy Observatory (NOAO), Southern Astrophysical Research (SOAR) 
and Gemini telescopes to build a network for rapid, flexible,
programmable access to world-class telescope facilities in the
forthcoming the LSST era (see Section \ref{sec:lsst}).

\vspace{0.1in}
Astronomers will be able to request observations from any
participating facility using the programmable and customizible
interface supported by the TOM Toolkit, as well as through the
facility's own system.  Telescopes in the network have agreed to some
or all of their time being available in highly flexible
queue-scheduling, so that observations can be requested at any time,
and conducted on timescales driven by the science goals.  The combined
network will offer access to imaging and spectroscopic instruments on
telescopes of a range of apertures distributed across the world. Each
facility retains control over the fraction of its time that is
executed in AEON mode, and telescope nightly operations need not be
fully robotic to participate.  Development for the successful first
phase of the AEON project was provided by the NSF through the NOAO
(e.g., $\simeq$ \$300K for the SOAR interfaces), with 
planning for the next phase currently underway.

\subsection{ {\bf \large \textcolor{CornflowerBlue}{Astropy - Kelle Cruz}} }\label{sec:astropy}

The Astropy Project is a community-driven effort to develop and
maintain a FOSS package that contains much of the core functionality and
common tools required across astronomy (\url{https://www.astropy.org}).  
These core functions include support for different file types, 
unit and physical quantity conversions, celestial coordinate and time transformations, 
world coordinate systems, containers for representing gridded
as well as tabular data, and a framework for cosmological
transformations and conversions
\citep{astropy-collaboration_2013_aa,astropy-collaboration_2018_aa}.
One goal is to avoid duplicated efforts for these common core tasks,
and to provide a robust framework upon which to build more complex
interoperable tools.  The Astropy Project also hosts an ecosystem of
affiliated packages that are not necessarily developed by the core
development team, but share the goals of Astropy, and often build from
the core package's code and infrastructure.

\vspace{0.1in}
Astropy has been extraordinarily well-received by the astronomy
community.  The first software article
\citep{astropy-collaboration_2013_aa} is currently in the 
\href{https://tinyurl.com/mesatwo2013}{Top 3}
most cited article in astronomy and astrophysics that
were published in 2013.  The second software article
\citep{astropy-collaboration_2018_aa} is presently also in the
\href{https://tinyurl.com/mesafour2018}{Top 3}
most cited astronomy and astrophysics articles that
were published in 2018.
The Astropy Project is also unique in having a very large, active team.  
Its unqualified success is made possible by the efforts of
$\simeq$ 200 team members that perform numerous important
roles. These roles encompass a broad scope of responsibilities ranging
from direct package development to communication, distribution, and
managerial activities.

\vspace{0.1in}
The Astropy Project has the ability to
accept financial contributions from institutions or individuals
through NumFOCUS, a 501(c)(3) public charity whose vision is an
inclusive scientific and research community that utilizes actively
supported open source software to make impactful discoveries for a
better world. The Astropy Project has recieved \$900k in funding from the Moore Foundation.

\subsection{ {\bf \large \textcolor{CornflowerBlue}{ATHENA++ - Jim Stone}} }\label{sec:athena}

Numerical methods are essential for exploring of a wide range of
applications in astrophysical fluid dynamics. The development
of accurate and capable algorithms, along with a description of their
implementation on modern large-scale parallel computer systems, is
crucial for progress in this field.  Athena++ is a software instrument
for performing such astrophysical magnetohydrodynamic simulations.
Athena++ itself is a complete re-write of the Athena software
instrument.  Currently, the instrument paper describing Athena
\citep{stone_2008_aa} ranks within the 
\href{https://tinyurl.com/u65orwz} {Top 60}
most cited papers ($\simeq$ 500 citations, $\simeq$ 60 citations/year) in astronomy 
that were published in 2008.

\vspace{0.1in}
Athena++ is an example of application scientists developing software
to solve science problems (see Section \ref{sec:gw} for another
example).  It is developed using continuous integration tools (e.g.,
Jenkins, Travis CI) and maintains strict adherence to C++ standards.
As a result, Athena++ has been run on everything from a laptop to the
very largest machines \citep[e.g.,][]{white_2016_aa,felker_2018_aa}.
The package is currently developed by 4 core team members, who
collectively do $\simeq$ 90\% of the commits, and about 8 significant
contributors (\url{https://princetonuniversity.github.io/athena/}). 
While the package is not intended to provide a service to the
community, the first user meeting was held in April 2019 at the
University of Nevada, Las Vegas and attended by $\simeq$ 60 (mostly young) 
participants interested in computational fluid dynamics.  Athena++ has
a recent download rate of $\simeq$ 300 time per month by its user
base.  Development and maintenance of Athena++ has been funded by
science grants. To date, there has been no funding for code
development or professional support for its $\simeq$ 100,000 lines of
code.  Athena++ will likely survive as a software instrument only as
long as it remains a useful tool for doing science problems.

\subsection{ {\bf \large \textcolor{CornflowerBlue}{Einstein Toolkit - Philipp M\"{o}sta}} }\label{sec:einstein}

Scientific progress in the field of numerical relativity has always
been closely tied to the availability and ease-of-use of enabling
software and computational infrastructure.
The Einstein Toolkit is a community-driven FOSS platform of core
computational tools to advance research in relativistic astrophysics
and gravitational physics (\url{https://einsteintoolkit.org}).  
The aim is to provide the core computational tools that can enable new
science, broaden the community, facilitate interdisciplinary research,
and take advantage of emerging petascale computers and advanced
digital infrastructure.  A large portion of the current Toolkit is
made up of over 100 community-developed components for computational
relativity along with associated tools for simulation management and
visualization.  The software instrument paper describing the Einstein
Toolkit \citep{loffler_2012_aa} currently has $\simeq$ 230 citations
and a citation rate of $\simeq$ 40/year, with individual components
that the Toolkit having their own bibliometrics.

\vspace{0.1in}
The Einstein Toolkit is currently used by $\simeq$ 240 users in
$\simeq$ 50 numerical relativity groups in the USA and Europe, with
most of users coming from $\simeq$ 10 research groups.  These
researchers use the Toolkit framework and generally deploy their
own proprietary components for applications.  Many of these users do
not directly collaborate on science problems, and in some cases
compete.  However, these groups agree that sharing the development of
the underlying infrastructure is mutually beneficial for every group
and the wider community as well.  The Toolkit does not itself develop
software. A distributed developer model ($\simeq$ 350 commits by $\simeq$ 20
team members within the past year) maintains core support of the
Toolkit with partnerships to $\simeq$ 7
core developers who contribute and coordinate together on
development. Official stable releases occur every 6 months.

\vspace{0.1in}
Funding for the development and maintenance of the Einstein Toolkit
include a NSF Physics at the Information Frontier Grant (2006-2015) to
Georgia Institute of Technology, California Institute of Technology,
Louisiana State University, and Rochester Institute of Technology at
$\simeq$~\$160K/yr/institution.  Current funding includes a NASA
Theoretical and Computational Astrophysics Network centered at
Rochester Institute of Technology ($\simeq$ \$1.6M between 2018-2021).
Current challenges identified include finding the balance between
curation and innovation, longevity and sustainability on decadal
timescales, and staying relevant on modern, evolving High Performance
Computing infrastructures.

\subsection{ {\bf \large \textcolor{CornflowerBlue}{PyCBC - Duncan Brown}} }\label{sec:gw}

The LIGO and Virgo detectors are large-scale experiments designed to directly detect
gravitational waves, from, for example, merging black holes
\citep[e.g.,][]{abbott_2016_ab}.  LIGO is large-scale NSF funded project 
involving thousands of people and hundreds of institutions.
The Gravitational Wave Open Science
Center (\url{https://www.gw-openscience.org/about/})  provides
calibrated data from these observatories, along with access to
tutorials and software tools \citep{vallisneri_2015_aa}.

\vspace{0.1in}
PyCBC is one of those FOSS software instruments (\url{https://pycbc.org}).  
PyCBC's algorithms can detect coalescing compact
binaries and provide Bayesian estimates of the astrophysical
parameters of detected sources \citep[e.g.,][]{nitz_2017_aa,biwer_2019_aa}.  
The software enables accessing the open gravitational wave data through the
Open Science Grid (OSG, \url{https://opensciencegrid.org}) and the
Extreme Science and Engineering Discovery Environment (XSEDE, \url{https://www.xsede.org}) 
computational platforms.  PyCBC was used in the first direct detection
of gravitational waves by LIGO and is used in the ongoing analysis of
LIGO and Virgo data. PyCBC was featured in Physics World \citep{smith_2017_aa}
as a good example of a large collaboration of application scientists developing
software to solve science problems that published its research
products and its software.
PyCBC has been mentioned in 186 articles since 2014, including five
ranked as top-cite 1000+, four ranked top-cite 500+,
and nine articles with over 100 citations (data from INSPIRE-HEP). The binary-merger search
method described in \citet{usman_2016_aa} has over 160 citations.

\vspace{0.1in}
PyCBC is developed on GitHub by about 6 core developers and about 70
contributors (roughly 90\% male, 10\% female) who directly support a community
of $\simeq$ 200 users.  Communication between developers and users
mainly occurs through Slack, which also serves as a repository of the
discussions with over 160,000 messages.  PyCBC's funding model for
development and maintenance is mainly through the science that it
enables, although there has been support through the NSF
Advanced Cyberinfrastructure program (\$500K between 2014 - 2019).  
The core science funding is through the NSF Gravity program (\$900K between 2014-2019), 
the Albert Einstein Institute (\$1M between 2014-2019), and awards 
to United Kingdom and European Union investigators.

\subsection{ {\bf \large \textcolor{CornflowerBlue}{JETFIT  - Andrew MacFadyen}} }\label{sec:jetfit}

Jets are collimated beams of matter ejected from an astronomical
object.  Relativistic jets occur when the beams of matter are
accelerated close to the speed of light.  Examples of objects that
feature high-energy jets include supermassive black holes in the
center of galaxies, X-ray binaries, gamma-ray bursts, and
protostars. JETFIT is a software tool to fit observational data of
jets to numerical simulations.  JETFIT is an example of application
scientists developing software to solve curiosity-driven science
problems.  It has a community of 2 or 3 researchers who constitute
both the developers and the users of this FOSS. It is hosted on
GitHub, and there is no help desk, community forum, outreach campaign,
or email listserv.  Science-driven development of JETFIT and the RAM
(Relativistic Adaptive Mesh) hydrodynamics package has been assisted,
at times, by science funding from the NASA SWIFT mission (\$15K), the
NASA NASA Astrophysics Theory Program (\$300K for 3 years), and the
NSF (\$300K for 3 years).

\subsection{ {\bf \large \textcolor{CornflowerBlue}{Astrophysics Source Code Library (ASCL) - Alice Allen}} }\label{sec:ascl}

The ASCL is an online registry of scientist-written software used in
astronomy or astrophysics research (\url{https://ascl.net}). The primary
objectives of the ASCL are to make the software used in research more
discoverable, more available, and more transparent
\citep{allen_2015_aa,allen_2018_aa,shamir_2018_aa}. Entries in the
ASCL include the name, description, author of the software, a unique
ASCL ID, and either a link to a download site for the software or an
attached archive file for the software so the code can be downloaded
directly from the ASCL. A link to an article describing or using the
software is usually included to demonstrate that the software has been
used for research in the refereed literature.  ASCL entries are
indexed by the SAO/NASA Astrophysics Data System (ADS) and Web of
Science's Data Citation Index.  Software can thus be cited in a
journal article even when there is no citable article describing the
software.  Additionally, ADS can link some articles which use software
to the software entries, enabling an easier examination of the
computational methods used. ADS also tracks citations to ASCL entries,
assuming the citations are suitably formatted, which can help authors
of research software for whom citations are an important metric.

\vspace{0.1in}
Started in 1999, ASCL initially required software to be deposited.
Most software authors were reluctant to adopt this model, which
stagnated ASCL's growth. ASCL dropped this model in 2010, although
software deposits are still accepted, and transitioned to the current
model of providing download locations.  ASCL subsequently grew from
$\simeq$ 40 entries to about 2100 today.  The percentage of software
that have been cited grew from $\simeq$ 7\% in 2014 to $\simeq$ 34\%
in 2019.  Citations in the refereed literature to ASCL entries
increased by $\simeq$ 59\% between 2017 and 2018, while the number of
entries increased by $\simeq$ 35\%.  This suggests a growth in the use
of ASCL over and above the growth in the number of ASCL
entries \citep{allen_2018_aa}. Finally, author submissions grew by
$\simeq$ 31\% over the same time period.

\vspace{0.1in}
ASCL is currently supported by the Heidelberg Institute for
Theoretical Studies ($\simeq$ \$6K/year), the NASA Astrophysics Data
Analysis Program ($\simeq$ \$162K for two years), in-kind
contributions from Michigan Technological University for internet
hosting, and in-kind contributions from the University of Maryland
Libraries for DOI minting.

\subsection{ {\bf \large \textcolor{CornflowerBlue}{MESA - Frank Timmes}} }\label{sec:mesa}

The Modules for Experiments in Stellar Astrophysics (MESA) source code
is a set of FOSS modules for stellar astrophysics that can be used on
their own, or combined to solve the coupled equations governing 1D
stellar evolution.  The MESA Project's community is stellar
astronomers and astrophysicists. The MESA Project has grown from
$\simeq$ 100 users in 2012 through $\simeq$ 800 users in 2015 to over
1000 users world-wide. It currently accounts for about 1/2 of all
stellar models published in the modern literature.  The MESA Project
innovates by updating its community with an instrument article
describing major new software and science capabilities about every two
years.  MESA I \citep{paxton_2011_aa} is currently ranked within the
\href{https://tinyurl.com/mesaone}{Top 5}
most cited articles in astronomy and astrophysics that were
published in 2011.  MESA II \citep{paxton_2013_aa} is in the 
\href{https://tinyurl.com/mesatwo2013}{Top 10}
most cited articles published in 2013, MESA III \citep{paxton_2015_aa} is in the 
\href{https://tinyurl.com/mesathree2015}{Top 10} 
most cited articles published in 2015, and 
the recent MESA IV \citep{paxton_2018_aa} is in the 
\href{https://tinyurl.com/mesafour2018}{Top 25} 
most cited articles published in 2018.  The articles that cite MESA have
themselves generated $\simeq$~45,000 citations, yielding a
radius-of-influence of $\simeq$ 15, which suggests that MESA helps
generate articles that the broader astronomy community values.

\vspace{0.1in}
The MESA Project provides two portals to openly share knowledge.
MESA-Users offers $\simeq$~10,000 archived and searchable posts of
community discussions about stars. The website
\href{http://cococubed.asu.edu/mesa_market/}{http://mesastar.org} offers a
Zenodo backed portal to build provenance by sharing tools and
guidance. Currently there are $\simeq$~400 contributions by the
community, for the community.  A Software Development Kit builds the
instrument across a variety of Unix-based platforms.  The offshoot
MESA-Docker simplifies the run-time requirements with only minor
overhead from running in a container, and is useful for new users and
Windows users.  Another spinoff is MESA-WEB (\url{http://mesa-web.asu.edu}), 
a cloud-based resource for education that has served 4000+ models to $\simeq$ 600 unique
users in $\simeq$ 3 years of operation.

\vspace{0.1in}
The MESA Project is currently developed by a team of 16, of which 2
are women ($\simeq$ 12\%). The percentage of women developers is 
smaller ($\simeq$ 20\%) than  recent demographics of physics and astronomy theory
Ph.D graduates.  The MESA Summer School, which averages $\simeq$ 36\%
women participants (e.g.,\url{http://cococubed.asu.edu/mesa_summer_school_2019/}), 
offers a week of extensive hands-on labs to gain familiarity with MESA
software instruent and learn how to make better use of MESA in their
own research.  The Summer School cadre of instructors, TAs and
participants (now over 250) are creating their own MESA user
infrastructure at $\simeq$ 40 institutions around the world.  The MESA
Project focuses on supporting young scientists who are also skilled at
developing community software instruments to obtain high-profile
graduate fellowships (6), named postdoc fellowships (6), and
tenure-stream positions (5). The MESA Project has been supported by
funding from the NSF Software Infrastructure for Sustained Innovation
program with \$500K between 2013-1016, and \$2.1M between 2017-2021.

\section{ {\bf \Large  \textcolor{CornflowerBlue}{Synthesis}} }\label{sec:discuss}

\setlength{\epigraphwidth}{0.7\textwidth}
\epigraph{Figuring out how to support digital infrastructure may seem daunting, but there are plenty of reasons to see the road ahead as an opportunity.}
{\textit{Nadia Eghbal \\ Roads And Bridges: The Unseen Labor Behind Our Digital Infrastructure}}

\vspace{0.1in}
The projects described above span a range of 
science (from observations through data pipelines to simulations), 
development models (from one person, through small teams, to large distributed alliances),
user communities (from a few to thousands), and 
funding models (from volunteers to institutional science grants).
Nevertheless, it is important to state that this sample is limited and 
incomplete in terms of the science spanned and projects studied. 
Any opinions, findings, conclusions or recommendations expressed below
should be weighed against these intrinsic imperfections.

\vspace{0.1in}
Returning to the research questions posed in the Executive summary,

\vspace{-0.1in}
\begin{itemize}\tightitemz
\item What is the relationship between money and sustainability for
community-driven, open-knowledge software instruments that enable
transformative research in stellar astrophysics? 
\item At what points in a software instrument’s lifecycle does an injection of financial resources help or hurt? 
\item Are science driven software instruments sustainable for the long term, say the next 40 years?
\end{itemize}\tightitemz 

\vspace{-0.1in}
\noindent
a potential funder should consider its desire to nurture a culture of
free, open-source, open-knowledge software sharing in astronomy.
Investments should be in a potential funder's wheelhouse: a focus on
cutting-edge basic science, a committment to sustaining software instruments 
that accelerate discovery, a desire to facilitate the participation of many diverse researchers, 
and a desire to realize an opportunity to fundamentally transform astronomy.

\vspace{0.1in}
Determining how much money is useful at a given stage of an individual
FOSS project's lifecycle will nearly always be a challenge.  Supporting
both up-and-coming and well-established software instruments
requires a degree of patience and a willingness to take risks.  
One key to success is to ensure the relevant scientific community is at
the helm and fully engaged with the software project.  The success of
the projects should be based on their open-source, open-knowledge
principles, their capability to enable open-ended discovery, their
integration of diversity and inclusion considerations throughout a
project, and their capacity to allow new questions in astronomy that 
were unable to be addressed due to closed-source, closed-knowledge practices.

\vspace{0.1in}
It is often easier to fund innovation-driven research and much more
difficult to fund maintenance and community support.  For example,
software development efforts in stellar astrophysics are usually
contained within innovative research proposals. One common approach is to
describe the software development as a sub-aim of research focused on
astrophysical discovery.
It can be easy to overlook that developing community-driven software instruments
involves many tasks that are not related to new capabilities, 
but on improving the existing software and supporting the user community. 
These essential tasks may not be innovative or directly related to writing source code, 
but they are critical to improving the quality of the software instrument 
and to building a robust, productive relationship with users. These essential efforts include
updating the software instrument in response to continuous changes in hardware infrastructure;
refactoring the software to improve its usability, maintainability,
reliability, efficiency, correctness, robustness, extensibility or
interoperability; 
fixing bugs and writing associated test suites to ensure that these bugs don't reappear; 
writing and updating source code documentation and users tutorials; 
managing public releases; 
actively supporting the user community by promptly providing advice, best practices, and guidance;
managing new feature contributions from users; 
recruiting and nuturing new developers;
building and maintaining the business case for the software project;
helping core developers land permnant positions;
disseminating information of community interest such as job opportunities within the field and announcements; 
building the community through summer schools, workshops, or seminars;
managing project infrastructure including source code repositories, web sites, mailing lists, 
wikis, social media feeds, Docker containers, and software development tools; and 
being agile in the face of new experimental or observational scientific discoveries.
A thriving open-knowledge software project takes a village,
whose activities take time, commitment, and financial resources.

\vspace{0.1in}
Software projects being considered for a potential funder's support
should go through a rigorous internal review process and an external
peer-review process. Questions that a potential funder may want to see 
addressed when considering support of a software project include:

\begin{itemize}\tightitemz

\item What does the software do? What are its key capabilities and novel features? 

\item Who is the target audience for the software?

\item What is an estimate of the size of the market for the software? What is the software's current market penetration?

\item To what extent does the software fill a recognized need within a community to advance basic science research?

\item Are there other FOSS projects that have similar functionality? What makes the software different from others, what is unique? 

\item How many person-hours, funded and unfunded, have been invested to date to bring the software to its current state?

\item What is the source code's license? Does the software project demonstrate a sufficient commitment to free, open-source, and open knowledge principles?

\item What tangible metrics will be used to measure the success of the software project?

\item To what extent has the software project demonstrated agility to new science discoveries and/or the changing needs of its the user community?

\item Hardware changes. Compilers change. Libraries change. 
To what extent is the software adaptable to new technologies? 

\item To what extent are provenance, reproducibility, composability, 
and interoperability part of the software project's activities?

\item To what extent does the software project address diversity and inclusion within its user community, developers, and leadership?

\item Is there a demonstrated commitment to work with other software projects within the community's software ecosystem?

\item How long does it usually take to go from proof-of-concept prototyping of a new software element 
to dissemination into the user community?

\item Has the software project been funded in the past? What was the impact of that funding?

\item How does the software project define sustainability?

\item To what extent are the current resources no longer adequate?

\item Why should funders help this software project thrive? 

\end{itemize}\tightitemz

To report on its progress, each year the software project's principals
send a summary of its progress relative to what was planned.  Each
project comes with an appropriate set of metrics - adding new
capabilities to the software, refactoring the instrument to the ever-evolving hardware, community engagement,
achieving scientific research objectives, or publishing a certain
number of software instrument papers.  The funder should review the
progress made each year and towards the end of a funding phase,
determine whether to consider funding for another phase.

\vspace{0.1in}
In determining funding for initial or future phases, 
a funder should ask whether the software instrument
will still be relevant over the period of the project as the
landscape of astronomical research changes and evolves; whether the
research questions to be enabled by improvments to the software
instrument are likely to remain at the forefront of astronomy during
this period; whether the project leadership team and plan remain
strong; whether plans for software access, storage, archiving, and
dissemination continue to be on the leading edge; whether the project has
made substantial strides in terms of diversity and inclusion; and
whether there is continuing demand for the software instrument from the
relevant research community.

\vspace{0.1in}
As the figure on the cover page suggests, the builders of digital infrastructure 
in stellar astrophysics -- be it for telescopes, data
pipelines, or theoretical models -- are deeply interconnected to each
other and technically very talented. They have already built the core
platforms that power much of modern astronomy. They just need help to
keep the gears rolling so they can continue doing what they do best.

\clearpage
\setcounter{page}{1}
\bibliographystyle{nsf}

% \bibliography{sfdi}

\begin{thebibliography}{29}
\expandafter\ifx\csname natexlab\endcsname\relax\def\natexlab#1{#1}\fi

\bibitem[{{Abbott} {et~al.}(2016){Abbott}, {Abbott}, {Abbott}, {Abernathy},
  {Acernese}, {Ackley}, {Adams}, {Adams}, {Addesso}, {Adhikari}, \&
  et~al.}]{abbott_2016_ab}
{Abbott}, B.~P., {Abbott}, R., {Abbott}, T.~D., {Abernathy}, M.~R., {Acernese},
  F., {Ackley}, K., {Adams}, C., {Adams}, T., {Addesso}, P., {Adhikari}, R.~X.,
  et~al., 2016, \textit{{Astrophysical Implications of the Binary Black Hole
  Merger GW150914}}, \textit{\apjl}, {818}, L22

\bibitem[{{Allen} \& {Schmidt}(2015)}]{allen_2015_aa}
{Allen}, A., {Schmidt}, J., 2015, \textit{{Looking before Leaping: Creating a
  Software Registry}}, \textit{Journal of Open Research Software}, {3}, E15

\bibitem[{{Allen} {et~al.}(2018){Allen}, {Teuben}, \& {Ryan}}]{allen_2018_aa}
{Allen}, A., {Teuben}, P.~J., {Ryan}, P.~W., 2018,
  \textit{{Schroedinger{\textquoteright}s Code: A Preliminary Study on Research
  Source Code Availability and Link Persistence in Astrophysics}},
  \textit{\apjs}, {236}, 10

\bibitem[{{Astropy Collaboration} {et~al.}(2018){Astropy Collaboration},
  {Price-Whelan}, {Sip{\H{o}}cz}, \& {et al.}}]{astropy-collaboration_2018_aa}
{Astropy Collaboration}, {Price-Whelan}, A.~M., {Sip{\H{o}}cz}, B.~M., {et
  al.}, 2018, \textit{{The Astropy Project: Building an Open-science Project
  and Status of the v2.0 Core Package}}, \textit{\aj}, {156}, 123

\bibitem[{{Astropy Collaboration} {et~al.}(2013){Astropy Collaboration},
  {Robitaille}, {Tollerud}, \& {et al.}}]{astropy-collaboration_2013_aa}
{Astropy Collaboration}, {Robitaille}, T.~P., {Tollerud}, E.~J., {et al.},
  2013, \textit{{Astropy: A community Python package for astronomy}},
  \textit{\aap}, {558}, A33

\bibitem[{{Biwer} {et~al.}(2019){Biwer}, {Capano}, {De}, {Cabero}, {Brown},
  {Nitz}, \& {Raymond}}]{biwer_2019_aa}
{Biwer}, C.~M., {Capano}, C.~D., {De}, S., {Cabero}, M., {Brown}, D.~A.,
  {Nitz}, A.~H., {Raymond}, V., 2019, \textit{{PyCBC Inference: A Python-based
  Parameter Estimation Toolkit for Compact Binary Coalescence Signal}},
  \textit{\pasp}, {131}, 024503

\bibitem[{{Cherinka} {et~al.}(2019){Cherinka}, {Andrews},
  {S{\'a}nchez-Gallego}, {Brownstein}, {Argudo-Fern{\'a}ndez}, {Blanton},
  {Bundy}, {Jones}, {Masters}, {Law}, {Rowlands}, {Weijmans}, {Westfall}, \&
  {Yan}}]{cherinka_2019_aa}
{Cherinka}, B., {Andrews}, B.~H., {S{\'a}nchez-Gallego}, J., {Brownstein}, J.,
  {Argudo-Fern{\'a}ndez}, M., {Blanton}, M., {Bundy}, K., {Jones}, A.,
  {Masters}, K., {Law}, D.~R., {Rowlands}, K., {Weijmans}, A.-M., {Westfall},
  K., {Yan}, R., 2019, \textit{{Marvin: A Tool Kit for Streamlined Access and
  Visualization of the SDSS-IV MaNGA Data Set}}, \textit{\aj}, {158}, 74

\bibitem[{{Eghbal}(2016)}]{eghbal_2016_aa}
{Eghbal}, N., 2016, \textit{{Roads And Bridges: The Unseen Labor Behind Our
  Digital Infrastructure}}. Tech. rep., {Ford Foundation}

\bibitem[{{Felker} \& {Stone}(2018)}]{felker_2018_aa}
{Felker}, K.~G., {Stone}, J.~M., 2018, \textit{{A fourth-order accurate finite
  volume method for ideal MHD via upwind constrained transport}},
  \textit{Journal of Computational Physics}, {375}, 1365-1400

\bibitem[{{Foreman-Mackey} {et~al.}(2013){Foreman-Mackey}, {Hogg}, {Lang}, \&
  {Goodman}}]{foreman-mackey_2013_aa}
{Foreman-Mackey}, D., {Hogg}, D.~W., {Lang}, D., {Goodman}, J., 2013,
  \textit{{emcee: The MCMC Hammer}}, \textit{\pasp}, {125}, 306

\bibitem[{{L{\"o}ffler} {et~al.}(2012){L{\"o}ffler}, {Faber}, {Bentivegna},
  {Bode}, {Diener}, {Haas}, {Hinder}, {Mundim}, {Ott}, {Schnetter}, {Allen},
  {Campanelli}, \& {Laguna}}]{loffler_2012_aa}
{L{\"o}ffler}, F., {Faber}, J., {Bentivegna}, E., {Bode}, T., {Diener}, P.,
  {Haas}, R., {Hinder}, I., {Mundim}, B.~C., {Ott}, C.~D., {Schnetter}, E.,
  {Allen}, G., {Campanelli}, M., {Laguna}, P., 2012, \textit{{The Einstein
  Toolkit: a community computational infrastructure for relativistic
  astrophysics}}, \textit{Classical and Quantum Gravity}, {29}, 115001

\bibitem[{{National Research
  Council}(2010)}]{national-research-council_2010_aa}
{National Research Council}, 2010, \textit{{New Worlds, New Horizons}}. New
  Worlds, New Horizons National Academy Press, Washington DC, 2010

\bibitem[{{National Research
  Council}(2011)}]{national-research-council_2011_aa}
---, 2011, \textit{{Vision and Voyages for Planetary Science in the Decade
  2013-2022}}. Vision and Voyages for Planetary Science in the Decade
  2013-2022, National Academy Press, Washington DC, 2010

\bibitem[{{Nitz} {et~al.}(2017){Nitz}, {Dent}, {Dal Canton}, {Fairhurst}, \&
  {Brown}}]{nitz_2017_aa}
{Nitz}, A.~H., {Dent}, T., {Dal Canton}, T., {Fairhurst}, S., {Brown}, D.~A.,
  2017, \textit{{Detecting Binary Compact-object Mergers with Gravitational
  Waves: Understanding and Improving the Sensitivity of the PyCBC Search}},
  \textit{\apj}, {849}, 118

\bibitem[{{Paxton} {et~al.}(2011){Paxton}, {Bildsten}, {Dotter}, {Herwig},
  {Lesaffre}, \& {Timmes}}]{paxton_2011_aa}
{Paxton}, B., {Bildsten}, L., {Dotter}, A., {Herwig}, F., {Lesaffre}, P.,
  {Timmes}, F., 2011, \textit{{Modules for Experiments in Stellar Astrophysics
  (MESA)}}, \textit{\apjs}, {192}, 3

\bibitem[{{Paxton} {et~al.}(2013){Paxton}, {Cantiello}, {Arras}, {Bildsten},
  {Brown}, {Dotter}, {Mankovich}, {Montgomery}, {Stello}, {Timmes}, \&
  {Townsend}}]{paxton_2013_aa}
{Paxton}, B., {Cantiello}, M., {Arras}, P., {Bildsten}, L., {Brown}, E.~F.,
  {Dotter}, A., {Mankovich}, C., {Montgomery}, M.~H., {Stello}, D., {Timmes},
  F.~X., {Townsend}, R., 2013, \textit{{Modules for Experiments in Stellar
  Astrophysics (MESA): Planets, Oscillations, Rotation, and Massive Stars}},
  \textit{\apjs}, {208}, 4

\bibitem[{{Paxton} {et~al.}(2015){Paxton}, {Marchant}, {Schwab}, {Bauer},
  {Bildsten}, {Cantiello}, {Dessart}, {Farmer}, {Hu}, {Langer}, {Townsend},
  {Townsley}, \& {Timmes}}]{paxton_2015_aa}
{Paxton}, B., {Marchant}, P., {Schwab}, J., {Bauer}, E.~B., {Bildsten}, L.,
  {Cantiello}, M., {Dessart}, L., {Farmer}, R., {Hu}, H., {Langer}, N.,
  {Townsend}, R.~H.~D., {Townsley}, D.~M., {Timmes}, F.~X., 2015,
  \textit{{Modules for Experiments in Stellar Astrophysics (MESA): Binaries,
  Pulsations, and Explosions}}, \textit{\apjs}, {220}, 15

\bibitem[{{Paxton} {et~al.}(2018){Paxton}, {Schwab}, {Bauer}, {Bildsten},
  {Blinnikov}, {Duffell}, {Farmer}, {Goldberg}, {Marchant}, {Sorokina},
  {Thoul}, {Townsend}, \& {Timmes}}]{paxton_2018_aa}
{Paxton}, B., {Schwab}, J., {Bauer}, E.~B., {Bildsten}, L., {Blinnikov}, S.,
  {Duffell}, P., {Farmer}, R., {Goldberg}, J.~A., {Marchant}, P., {Sorokina},
  E., {Thoul}, A., {Townsend}, R.~H.~D., {Timmes}, F.~X., 2018,
  \textit{{Modules for Experiments in Stellar Astrophysics (MESA): Convective
  Boundaries, Element Diffusion, and Massive Star Explosions}}, \textit{\apjs},
  {234}, 34

\bibitem[{{Shamir} {et~al.}(2018){Shamir}, {Berriman}, {Teuben}, {Nemiroff}, \&
  {Allen}}]{shamir_2018_aa}
{Shamir}, L., {Berriman}, B., {Teuben}, P., {Nemiroff}, R., {Allen}, A., 2018,
  \textit{{Best Practices for a Future Open Code Policy: Experiences and Vision
  of the Astrophysics Source Code Library}}, \textit{arXiv e-prints},
  arXiv:1802.00552

\bibitem[{{Smith}(2017)}]{smith_2017_aa}
{Smith}, A., 2017, \textit{{Why we should give credit to code creators}},
  \textit{Physics World}, {30}, 38-41

\bibitem[{{Stone} {et~al.}(2008){Stone}, {Gardiner}, {Teuben}, {Hawley}, \&
  {Simon}}]{stone_2008_aa}
{Stone}, J.~M., {Gardiner}, T.~A., {Teuben}, P., {Hawley}, J.~F., {Simon},
  J.~B., 2008, \textit{{Athena: A New Code for Astrophysical MHD}},
  \textit{\apjs}, {178}, 137-177

\bibitem[{{Townsend} {et~al.}(2018){Townsend}, {Goldstein}, \&
  {Zweibel}}]{townsend_2018_aa}
{Townsend}, R.~H.~D., {Goldstein}, J., {Zweibel}, E.~G., 2018, \textit{{Angular
  momentum transport by heat-driven g-modes in slowly pulsating B stars}},
  \textit{\mnras}, {475}, 879-893

\bibitem[{{Townsend} \& {Teitler}(2013)}]{townsend_2013_aa}
{Townsend}, R.~H.~D., {Teitler}, S.~A., 2013, \textit{{GYRE: an open-source
  stellar oscillation code based on a new Magnus Multiple Shooting scheme}},
  \textit{\mnras}, {435}, 3406-3418

\bibitem[{{Turk}(2013)}]{turk_2013_aa}
{Turk}, M.~J., 2013, \textit{{How to Scale a Code in the Human Dimension}},
  \textit{ArXiv e-prints}

\bibitem[{{Turk} {et~al.}(2011){Turk}, {Smith}, {Oishi}, {Skory}, {Skillman},
  {Abel}, \& {Norman}}]{turk_2011_aa}
{Turk}, M.~J., {Smith}, B.~D., {Oishi}, J.~S., {Skory}, S., {Skillman}, S.~W.,
  {Abel}, T., {Norman}, M.~L., 2011, \textit{{yt: A Multi-code Analysis Toolkit
  for Astrophysical Simulation Data}}, \textit{\apjs}, {192}, 9

\bibitem[{{Usman} {et~al.}(2016){Usman}, {Nitz}, {Harry}, {Biwer}, {Brown},
  {Cabero}, {Capano}, {Dal Canton}, {Dent}, {Fairhurst}, {Kehl}, {Keppel},
  {Krishnan}, {Lenon}, {Lundgren}, {Nielsen}, {Pekowsky}, {Pfeiffer},
  {Saulson}, {West}, \& {Willis}}]{usman_2016_aa}
{Usman}, S.~A., {Nitz}, A.~H., {Harry}, I.~W., {Biwer}, C.~M., {Brown}, D.~A.,
  {Cabero}, M., {Capano}, C.~D., {Dal Canton}, T., {Dent}, T., {Fairhurst}, S.,
  {Kehl}, M.~S., {Keppel}, D., {Krishnan}, B., {Lenon}, A., {Lundgren}, A.,
  {Nielsen}, A.~B., {Pekowsky}, L.~P., {Pfeiffer}, H.~P., {Saulson}, P.~R.,
  {West}, M., {Willis}, J.~L., 2016, \textit{{The PyCBC search for
  gravitational waves from compact binary coalescence}}, \textit{Classical and
  Quantum Gravity}, {33}, 215004

\bibitem[{{Vallisneri} {et~al.}(2015){Vallisneri}, {Kanner}, {Williams},
  {Weinstein}, \& {Stephens}}]{vallisneri_2015_aa}
{Vallisneri}, M., {Kanner}, J., {Williams}, R., {Weinstein}, A., {Stephens},
  B., 2015, \textit{{The LIGO Open Science Center}}, in Journal of Physics
  Conference Series, Vol. 610, \textit{Journal of Physics Conference Series},
  012021

\bibitem[{{Weaver} {et~al.}(2015){Weaver}, {Blanton}, {Brinkmann},
  {Brownstein}, \& {Stauffer}}]{weaver_2015_aa}
{Weaver}, B.~A., {Blanton}, M.~R., {Brinkmann}, J., {Brownstein}, J.~R.,
  {Stauffer}, F., 2015, \textit{{The Sloan Digital Sky Survey Data Transfer
  Infrastructure}}, \textit{\pasp}, {127}, 397

\bibitem[{{White} {et~al.}(2016){White}, {Stone}, \& {Gammie}}]{white_2016_aa}
{White}, C.~J., {Stone}, J.~M., {Gammie}, C.~F., 2016, \textit{{An Extension of
  the Athena++ Code Framework for GRMHD Based on Advanced Riemann Solvers and
  Staggered-mesh Constrained Transport}}, \textit{\apjs}, {225}, 22

\end{thebibliography}

\end{document}